\documentclass[a4paper,11pt]{article}
\usepackage{pos}

\title{Observation of the galactic PeVatron candidate LHAASO J2108+5157 with the Large-Sized Telescope for Cherenkov Telescope Array}
\ShortTitle{Observation of LHAASO J2108+5157 with the LST for CTA}

\manuallySeparateAuthors
\author*[a]{J. Jury\v{s}ek,}
\author[a]{M. Balbo,}
\author[a]{D. Eckert,}
\author[a]{A. Tramacere,}
\author[b]{G. Pirola}
\author[a]{and R. Walter}
\author{for the CTA-LST project}
 
\affiliation[a]{Department of Astronomy, University of Geneva\\
    Chemin d'Ecogia 16, CH-1290 Versoix, Switzerland}
\affiliation[b]{Max-Planck-Institut f\"{u}r Physik,\\
    Föhringer Ring 6, 80805 M\"{u}nchen, Germany}

\emailAdd{lst-contact@cta-observatory.org}
\emailAdd{jakub.jurysek@unige.ch}

\abstract{The Cherenkov Telescope Array (CTA) Observatory will be the next generation ground-based very-high-energy gamma-ray observatory, sensitive from 20 GeV up to 300 TeV. The Large-Sized Telescope prototype (LST-1), currently in the commissioning phase, was inaugurated in October 2018 on La Palma (Spain). It is the first of four LST telescopes for CTA, to be built in La Palma. In 2021, LST-1 performed observations of one of the Galactic PeVatron candidates, LHAASO J2108+5157, recently discovered by the LHAASO collaboration. We present results of our analysis of the LST-1 data, putting strong constraints on the emission of the source in the multi-TeV band. We also present results of multi-wavelength modelling using 12-years Fermi-LAT data and Target of Opportunity observations with XMM-Newton. We test different scenarios for the parent particles producing the high energy emission and put constraints on their spectra.}

\FullConference{%
  7th Heidelberg International Symposium on High-Energy Gamma-Ray Astronomy (Gamma2022)\\
  4-8 July 2022\\
  Barcelona, Spain\\}

\begin{document}
\maketitle

\section{Introduction}

Galactic PeVatrons are enigmatic cosmic accelerators, believed to produce cosmic rays (CRs) with energies up to at least $E \approx 1 \, \mathrm{PeV}$. When accelerated CR protons collide with the surrounding ambient gas, they subsequently emit gamma rays via $\pi^0$ decay with the energy of the gamma-ray photons about 1/10 that of the incident relativistic protons. Therefore, careful morphological and spectral studies of Ultra-High-Energy (UHE) gamma-ray sources emitting radiations with energies beyond $100 \, \mathrm{TeV}$ are needed to shed some light on the puzzle of the origin of Very-High-Energy (VHE) Galactic CRs (for a review, see e.g. \cite{2019IJMPD..2830022G}). However, the presence of UHE gamma-ray emission is not evidence of a hadronic accelerator, since gamma rays can also be produced by electrons and positrons via inverse Compton (IC) scattering on low-energy photon fields, or via bremsstrahlung on atomic nuclei of surrounding matter. They can also emit synchrotron radiation when travelling across a magnetic field. Although the IC emission of gamma rays with energies greater than $\sim 100 \, \mathrm{TeV}$ is suppressed due to the Klein-Nishina effect, IC can still dominate UHE emission in radiation-dominated environments \cite[e.g.][]{2009A&A...497...17V, 2021ApJ...908L..49B}. The nature of PeVatrons thus remains unknown despite substantial observational efforts in the last decades, when several UHE gamma-ray sources have been discovered with ground-based gamma-ray observatories, like Tibet AS$\gamma$ \cite{2019PhRvL.123e1101A, 2021NatAs...5..460T}, High Altitude Water Cherenkov (HAWC) Observatory \cite{2020PhRvL.124b1102A} and Large High Altitude Air Shower Observatory (LHAASO) \cite{2021Natur.594...33C}.

LHAASO J2108+5157 is the first gamma-ray source discovered in the UHE band, with no VHE counterpart known beforehand \cite{2021ApJ...919L..22C}. There is also no X-ray counterpart to the source, but a close High-Energy (HE) soft point source 4FGL J2108.0+5155 was identified at angular distance of $0.13^\circ$ \cite{2021ApJ...919L..22C}. There are also two molecular clouds in the direction coincident with LHAASO J2108+5157, which would support the hypothesis of hadronic origin of the emission, if CR protons collide with the ambient gas and emit gamma rays via $\pi^0$ decay. Leptonic emission, however, cannot be a priory excluded. The $95\%$ confidence level upper limit (UL) on the UHE source extension is $0.26^\circ$ \cite{2021ApJ...919L..22C} .

In this contribution, we present results of a dedicated observation of the source with the Large-Sized Telescope prototype (LST-1), results of Target of Opportunity observations with XMM-Newton, and also results of binned analysis of Fermi-LAT data, which together provide strong constraints on gamma-ray and X-ray emission of the source. We also present a modeling of multi-wavelength emission of the source and discuss possible emission scenarios.  This contribution is based on a study of the LST Collaboration \cite{2022arXiv221000775A}, where one can find further details on data analysis, results and modeling of the source emission.

\section{Observation and data analysis}

\subsection{LST-1}

LST-1 is the first single mirror dish (23 m in diameter) Imaging Air Cherenkov Telescope prototype built for the future Cherenkov Telescope Array Observatory (CTAO)\footnote{\url{https://www.cta-observatory.org/}} in La Palma island, Spain \cite{2021arXiv210806005M}. In single-telescope regime it is most sensitive in the energy range between tens of GeV and $\approx5$ TeV, with the energy and angular resolution of $20\,\%$ and $0.15^\circ$, respectively \cite{CTALSTproject:2021mfp}, [LST Collaboration, in prep.].

We observed LHAASO J2108+5157 with LST-1 for 91 hours in 2021. After application of quality cuts based on trigger rate stability, rate of CR events and atmospheric transmission, we were left with 49.3 hours of good quality data for the further analysis. For data calibration and event reconstruction we used a pipeline implemented in \texttt{lstchain v0.9.6.} \cite{ruben_lopez_coto_2022_6458862}. We analyzed $\theta^2$ distributions using reflected background method in four logarithmically spaced energy bins in the energy range between 100 GeV and 100 TeV, which resulted in no significant source detection. At the highest energies $E > 3 \, \mathrm{TeV}$, however, we see a point-like excess with a significance of $3.7\sigma$ \cite{1983ApJ...272..317L}. 

As a next step, we performed 1D spectral analysis with the use of the \texttt{gammapy} package \cite{2017ICRC...35..766D} in full LST-1 energy range. We assumed power law (PL) spectral shape $\mathrm{d}N/\mathrm{d}E = N_0 (E/E_0)^{-\Gamma}$ for the source spectrum. The reference energy was fixed as $E_0 = 1$ TeV. The best fit spectral parameters are $N_0 = (8.0 \pm 5.4) \times 10^{-14} \, \mathrm{cm^{-2} s^{-1} TeV^{-1}}$ and $\Gamma = 1.6 \pm 0.2$, suggesting hard emission of the source in the TeV band. To estimate source flux in individual energy bins, we run a joint likelihood fit of the LST-1 data and LHAASO flux-points, considering spectral shape in a form of power-law with exponential cutoff. Even though significance of a point-like source in the full LST-1 energy range is not high enough to claim a detection ($2.2\sigma$ for PL spectral model), LST-1 data provide strong ULs on the source emission in the TeV band (see Figure~\ref{fig_models}). 

\subsection{Fermi-LAT}
We performed a dedicated binned analysis of the 12-year Fermi-LAT data for the region. In contrast with the Cao et al. \cite{2021ApJ...919L..22C}, that used 4FGL-DR2 catalogue, we used more recent 12-years 4FGL-DR3 catalogue \cite{2022arXiv220111184F} to create the source model. The spectrum of LHAASO J2108+5157 counterpart, 4FGL J2108.0+5155, presents steep decrease above few GeVs and its link to the VHE-UHE emission is therefore not straightforward (see Figure~\ref{fig_models}). We also found that the emission above $\approx4$ GeV is dominated by a new hard source ($\Gamma = 1.9 \pm 0.2$), located at coordinates $l = 92.35^\circ$, $b = 2.56^\circ$, detected with a significance of $4\sigma$. This new source is not spatially correlated with LHAASO J2108+5157, but including it in the background model we were able to improve spectral representation of the HE counterpart 4FGL J2108.0+5155.

\subsection{XMM-Newton}

Target of Opportunity XMM-Newton observation, centered at $\mathrm{R.A.}= 317.0170^\circ$, $\mathrm{Dec}= +51.9275^\circ$, was carried out on June 11, 2021, collecting 13.6 ks of data. We reduced the data from all three (MOS1, MOS2, PN) EPIC cameras using \texttt{XMMSAS v19.1} and the \texttt{X-COP} data analysis pipeline \cite{eckert17,ghirardini19}. We extracted images in the soft (0.5-2 keV) and hard (2-7 keV) bands. The brightest X-ray source in the field is the eclipsing binary V1061 Cyg (RX J2107.3+5202), which exhibits very soft thermal spectrum and thus cannot be a counterpart of LHAASO J2108+5157. The other sources in the field are substantialy fainter. 

\begin{figure*}
\centering
%\resizebox{\hsize}{!}
        {
        \includegraphics[width=0.47\hsize]{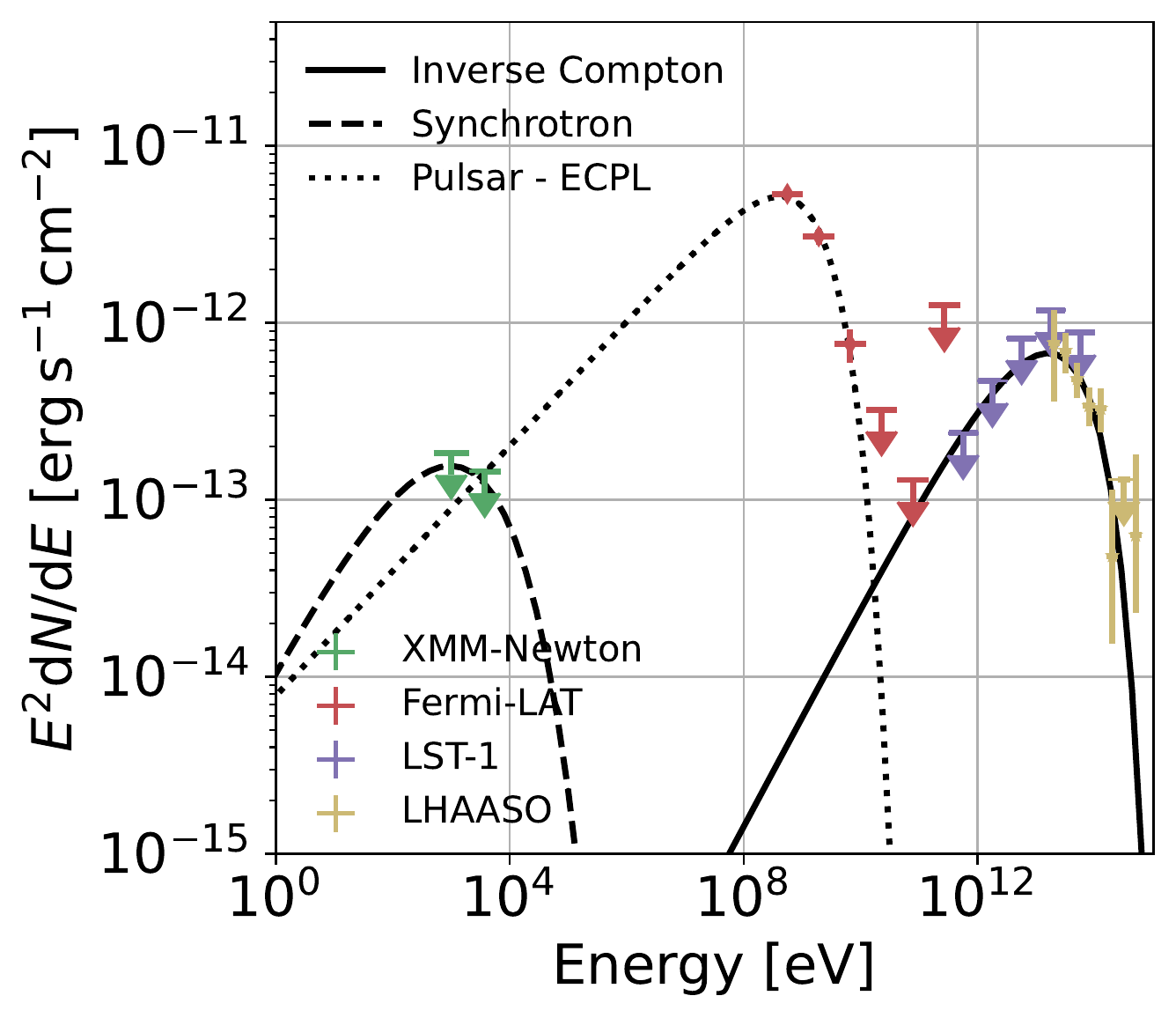}
        \includegraphics[width=0.50\hsize]{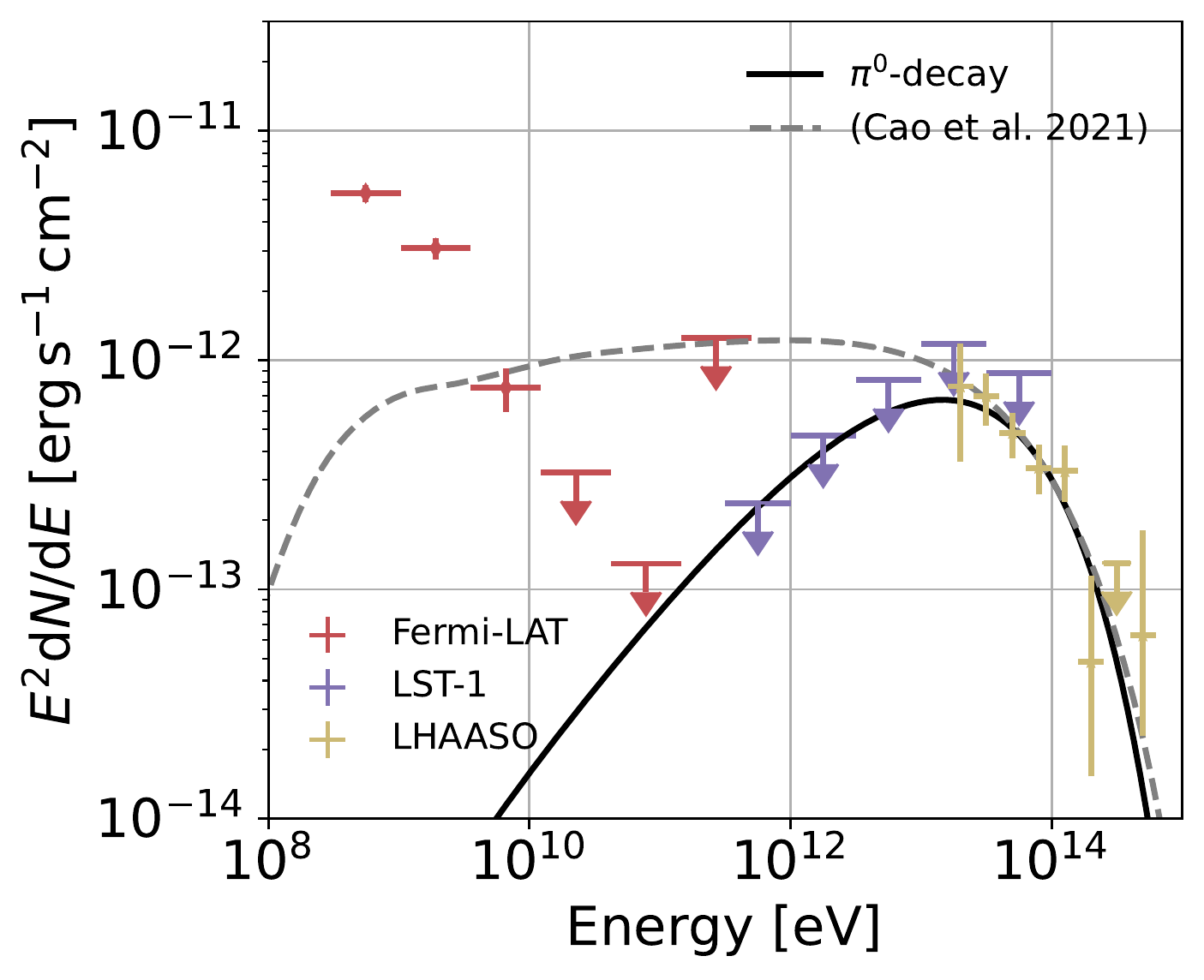}
        }
  \caption{\textit{Left:} Spectral Energy Distribution (SED) of LHAASO J2108+5157 showing leptonic emission scenario. Data from different instruments are shown - XMM-Newton (green), Fermi-LAT (red), LST-1 (purple), LHAASO-KM2A (yellow). The solid line represents the best fitting IC emission of LST-1 and LHAASO data and dashed line represents the synchrotron radiation of the same electrons in magnetic field of $B = 1.2 \, \mathrm{\mu G}$. Dotted line represents a phenomenological model of the tentative pulsar. \textit{Right:} SED with hadronic models of emission. The solid line represents the best fitting $\pi^0$ decay emission model. Grey dashed line represents $\pi^0$ decay emission model with proton spectral index $\alpha = 2$ shown for reference \cite{2021ApJ...919L..22C}.}
     \label{fig_models}
\end{figure*}

In the case of a leptonic origin of the source emission, we would expect a synchrotron nebula with an angular size of a few arcminutes, spatially correlated with the extended UHE emission. We used the public code \texttt{pyproffit} \cite{eckert20} to derive ULs on the X-ray flux of an extended source centered on the position of LHAASO J2108+5157 assuming PL source spectrum with a photon index of 2. Provided the unknown distance to the source, we derived ULs for two extreme cases - completely unabsorbed and completely absorbed source, for the source being located nearby, or on the other side of the Galaxy, respectively. Figure~\ref{fig_models} shows $95\%$ absorbed ULs on the X-ray emission in a region with radius of $6'$, which is a maximum reasonable X-ray extension of a Galactic Pulsar Wind Nebula (PWN) at a distance $d \leq 10$ kpc \cite{Bamba_2010}.
%In order to convert ULs on detected counts to fluxes, one has to take into account the effect of Galactic extinction, which is substantial along the Galactic plane - HI column density at the position of the LHAASO source is $1.05 \times 10^{22} \, \mathrm{cm^{-2}}$ \cite{2016A&A...594A.116H}. 

\section{Discussion of possible emission scenarios}

\subsection{Leptonic emission scenario}

The fact that most of the currently known VHE sources have been identified as PWN \cite{2018A&A...612A...1H} and most of the potential Galactic PeVatrons can be associated with a nearby pulsar \cite{2021Natur.594...33C} favours a leptonic emission mechanism. We performed leptonic modelling of the multi-TeV emission in \texttt{naima} package \cite{2015ICRC...34..922Z}, assuming exponential cutoff power law distribution of electrons interacting via IC with Cosmic Microwave Background (CMB) and Far Infrared Radiation of the dust (FIR). The electron distribution best fitting the observations shown in Figure~\ref{fig_models} has a cutoff at $E_\mathrm{cutoff} = 100^{+70}_{-30}$ TeV and spectral index $\alpha = 1.5 \pm 0.4$. The total energy in electrons is $E(>1 \, \mathrm{GeV}) \approx 1 \times 10^{45} (d / 1 \, \mathrm{kpc)^2 \, erg}$.  X-ray ULs on synchrotron nebula emission strongly constrain magnetic field in the source ($B \leq 1.2 \, \mathrm{\mu G}$), which is hard to explain in PWN scenario, but consistent with TeV halo object \cite[e.g.][]{2022NatAs...6..199L}, where one can expect magnetic field even below Galactic background level \cite{Liu_2019}. 

The soft GeV emission of 4FGL J2108.0+5155 has spectral features compatible with a population of gamma-ray pulsars with spin-down power in the range of $\dot{E} \approx 10^{35} - 10^{37} \, \mathrm{erg \, s^{-1}}$ \cite{2013ApJS..208...17A}. Cooling time of relativistic electrons of the cutoff energy is $t_\mathrm{cool} \approx 30$ kyr and thus such a pulsar could release enough energy in $t_\mathrm{cool}$ to drive the UHE emission of the TeV halo, provided the distance to the source $d \leq 10$ kpc. The possible extent of the UHE source is also consistent with the size of a typical TeV halo object, assuming Geminga-like diffusion coefficient and source distance $d \geq 2$ kpc.

\subsection{Hadronic emission scenario}

We created a model of $\pi^0$ decay dominated multi-TeV emission in \texttt{naima} package, assuming relativistic protons following exponential cutoff power law distribution, interacting with one of the two molecular clouds in the direction coincident with LHAASO J2108+5157. The best fitting proton distribution shown in Figure~\ref{fig_models} has cutoff at $E_\mathrm{cutoff} = 250^{+200}_{-90}$ TeV and spectral index $\alpha = 0.8 \pm 0.5$. While the total energy in relativistic protons is low enough ($E_\mathrm{total} = 9 \times 10^{45} (d / 1 \, \mathrm{kpc})^2 (100 \, \mathrm{cm^{-3}} / n) \, \mathrm{erg}$, where $n$ is molecular density of the molecular cloud with distance $d$ in the direction of the source) compared to the typical energy of a supernova explosion, required spectral index is too hard compared to the standard diffusive shock acceleration (DSA) \cite{1978MNRAS.182..147B}. However, it does not necessarily exclude hadronic emission scenario as hard spectral indices can be expected e.g. in the non-linear regime of DSA \cite{Berezhko_1999}.

The HE gamma-ray emission of the Fermi-LAT counterpart cannot be explained in a single component model together with UHE emission in hadronic scenario. Cao et al. \cite{2021ApJ...919L..22C} suggested that 4FGL J2108.0+5155 show spectral features of an old SNR. Our dedicated analysis of Fermi-LAT data, however, results in the photon index of HE emission $\Gamma = -3.2$ above 1 GeV, which seems to be too soft compared to the spectra of old SNRs interacting with molecular clouds \cite{2012ApJ...761..133Y}.

\section{Conclusions}
We conducted a multi-wavelength study of the unidentified UHE gamma-ray source LHAASO J2108+5157 combining data from LST-1, XMM-Newton, Fermi-LAT and LHAASO. Strong ULs on emission of the source derived from the LST-1 data prove that LST-1 is capable of providing significant scientific results although still in commissioning phase. We present a leptonic emission scenario suggesting that the UHE source may be related to a TeV halo object. There is only a handful of TeV halo objects currently known, and these objects are of a particular interest because they appear to have particle diffusion coefficients that are much lower than in the interstellar medium, and thus may provide important insight into CR particle diffusion within the Galaxy. Soft HE counterpart can be explained as a pulsar, which could provide the relativistic electrons with enough energy to power the UHE emission. In the hadronic emission scenario, on the other hand, the HE and UHE emission cannot be explained as interaction of protons accelerated in an old SNR with dense molecular cloud, and thus more complex model would be needed.
%The lack of detection of a pulsed emission associated with the UHE source, however, presents a challenge for the TeV halo scenario.

%The HE gamma-ray emission in the hadronic scenario cannot be explained in a single component model together with UHE emission, which presents a challenge for the hadronic emission model.

\acknowledgments

We gratefully acknowledge financial support from the following agencies and organisations listed here \url{https://www.lst1.iac.es/acknowledgements.html}

This work was conducted in the context of the CTA-LST Project. This research has made use of the observations obtained with \emph{XMM-Newton}, an ESA science mission with instruments and contributions directly funded by ESA Member States and NASA. This research has made use of the SIMBAD database, operated at CDS, Strasbourg, France.

\bibliographystyle{JHEPe}
{\footnotesize
\bibliography{references}}

\providecommand{\href}[2]{#2}\begingroup\raggedright\begin{thebibliography}{10}

\bibitem{2019IJMPD..2830022G}
S.~{Gabici}, C.~{Evoli}, D.~{Gaggero}, P.~{Lipari}, P.~{Mertsch}, E.~{Orlando}
  et~al., \href{https://doi.org/10.1142/S0218271819300222}{\emph{International
  Journal of Modern Physics D} {\bfseries 28} (2019) 1930022}
  [\href{https://arxiv.org/abs/1903.11584}{{\ttfamily 1903.11584}}].

\bibitem{2009A&A...497...17V}
G.~{Vannoni}, S.~{Gabici} and F.~A. {Aharonian},
  \href{https://doi.org/10.1051/0004-6361/200809744}{\emph{A\&A} {\bfseries
  497} (2009) 17} [\href{https://arxiv.org/abs/0803.1138}{{\ttfamily
  0803.1138}}].

\bibitem{2021ApJ...908L..49B}
M.~{Breuhaus}, J.~{Hahn}, C.~{Romoli}, B.~{Reville}, G.~{Giacinti}, R.~{Tuffs}
  et~al., \href{https://doi.org/10.3847/2041-8213/abe41a}{\emph{ApJ} {\bfseries
  908} (2021) L49} [\href{https://arxiv.org/abs/2010.13960}{{\ttfamily
  2010.13960}}].

\bibitem{2019PhRvL.123e1101A}
M.~{Amenomori}, Y.~W. {Bao}, X.~J. {Bi}, D.~{Chen}, T.~L. {Chen}, W.~Y. {Chen}
  et~al.,
  \href{https://doi.org/10.1103/PhysRevLett.123.051101}{\emph{Phys.~Rev.~Lett.}
  {\bfseries 123} (2019) 051101}
  [\href{https://arxiv.org/abs/1906.05521}{{\ttfamily 1906.05521}}].

\bibitem{2021NatAs...5..460T}
{Tibet AS{\ensuremath{\gamma}} Collaboration}, M.~{Amenomori}, Y.~W. {Bao},
  X.~J. {Bi}, D.~{Chen}, T.~L. {Chen} et~al.,
  \href{https://doi.org/10.1038/s41550-020-01294-9}{\emph{Nature Astronomy}
  {\bfseries 5} (2021) 460} [\href{https://arxiv.org/abs/2109.02898}{{\ttfamily
  2109.02898}}].

\bibitem{2020PhRvL.124b1102A}
A.~U. {Abeysekara}, A.~{Albert}, R.~{Alfaro}, J.~R. {Angeles Camacho}, J.~C.
  {Arteaga-Vel{\'a}zquez}, K.~P. {Arunbabu} et~al.,
  \href{https://doi.org/10.1103/PhysRevLett.124.021102}{\emph{Phys.~Rev.~Lett.}
  {\bfseries 124} (2020) 021102}
  [\href{https://arxiv.org/abs/1909.08609}{{\ttfamily 1909.08609}}].

\bibitem{2021Natur.594...33C}
Z.~{Cao}, F.~A. {Aharonian}, Q.~{An}, L.~X. {Axikegu}, Bai, Y.~X. {Bai}, Y.~W.
  {Bao} et~al., \href{https://doi.org/10.1038/s41586-021-03498-z}{\emph{Nature}
  {\bfseries 594} (2021) 33}.

\bibitem{2021ApJ...919L..22C}
Z.~{Cao}, F.~{Aharonian}, Q.~{An}, {Axikegu}, L.~X. {Bai}, Y.~X. {Bai} et~al.,
  \href{https://doi.org/10.3847/2041-8213/ac2579}{\emph{ApJl} {\bfseries 919}
  (2021) L22} [\href{https://arxiv.org/abs/2106.09865}{{\ttfamily
  2106.09865}}].

\bibitem{2022arXiv221000775A}
S.~{Abe}, A.~{Aguasca-Cabot}, I.~{Agudo}, N.~{Alvarez Crespo}, L.~A.
  {Antonelli}, C.~{Aramo} et~al., {\emph{arXiv e-prints} (2022)
  arXiv:2210.00775} [\href{https://arxiv.org/abs/2210.00775}{{\ttfamily
  2210.00775}}].

\bibitem{2021arXiv210806005M}
D.~{Mazin}, {\emph{arXiv e-prints} (2021) arXiv:2108.06005}
  [\href{https://arxiv.org/abs/2108.06005}{{\ttfamily 2108.06005}}].

\bibitem{CTALSTproject:2021mfp}
{\scshape LST} collaboration,  in \emph{{37th International Cosmic Ray
  Conference}}, 9, 2021, \href{https://arxiv.org/abs/2109.03515}{{\ttfamily
  2109.03515}}.

\bibitem{ruben_lopez_coto_2022_6458862}
R.~Lopez-Coto, T.~Vuillaume, A.~Moralejo, F.~Cassol, M.~Nöthe, D.~Morcuende
  et~al.,  Apr., 2022.
\newblock 10.5281/zenodo.6458862.

\bibitem{1983ApJ...272..317L}
T.~P. {Li} and Y.~Q. {Ma}, \href{https://doi.org/10.1086/161295}{\emph{ApJ}
  {\bfseries 272} (1983) 317}.

\bibitem{2017ICRC...35..766D}
C.~{Deil}, R.~{Zanin}, J.~{Lefaucheur}, C.~{Boisson}, B.~{Khelifi},
  R.~{Terrier} et~al.,  in \emph{35th International Cosmic Ray Conference
  (ICRC2017)}, vol.~301 of \emph{International Cosmic Ray Conference}, p.~766,
  Jan., 2017, \href{https://arxiv.org/abs/1709.01751}{{\ttfamily 1709.01751}}.

\bibitem{2022arXiv220111184F}
{Fermi-LAT collaboration}, {:}, S.~{Abdollahi}, F.~{Acero}, L.~{Baldini},
  J.~{Ballet} et~al., {\emph{arXiv e-prints} (2022) arXiv:2201.11184}
  [\href{https://arxiv.org/abs/2201.11184}{{\ttfamily 2201.11184}}].

\bibitem{eckert17}
D.~{Eckert}, S.~{Ettori}, E.~{Pointecouteau}, S.~{Molendi}, S.~{Paltani} and
  C.~{Tchernin},
  \href{https://doi.org/10.1002/asna.201713345}{\emph{Astronomische
  Nachrichten} {\bfseries 338} (2017) 293}
  [\href{https://arxiv.org/abs/1611.05051}{{\ttfamily 1611.05051}}].

\bibitem{ghirardini19}
V.~{Ghirardini}, D.~{Eckert}, S.~{Ettori}, E.~{Pointecouteau}, S.~{Molendi},
  M.~{Gaspari} et~al.,
  \href{https://doi.org/10.1051/0004-6361/201833325}{\emph{A\&A} {\bfseries
  621} (2019) A41} [\href{https://arxiv.org/abs/1805.00042}{{\ttfamily
  1805.00042}}].

\bibitem{eckert20}
D.~{Eckert}, A.~{Finoguenov}, V.~{Ghirardini}, S.~{Grandis}, F.~{Kaefer},
  J.~{Sanders} et~al.,
  \href{https://doi.org/10.21105/astro.2009.03944}{\emph{The Open Journal of
  Astrophysics} {\bfseries 3} (2020) 12}
  [\href{https://arxiv.org/abs/2009.03944}{{\ttfamily 2009.03944}}].

\bibitem{Bamba_2010}
A.~Bamba, T.~Anada, T.~Dotani, K.~Mori, R.~Yamazaki, K.~Ebisawa et~al.,
  \href{https://doi.org/10.1088/2041-8205/719/2/l116}{\emph{The Astrophysical
  Journal} {\bfseries 719} (2010) L116}.

\bibitem{2018A&A...612A...1H}
{H.~E.~S.~S. Collaboration}, H.~{Abdalla}, A.~{Abramowski}, F.~{Aharonian},
  F.~{Ait Benkhali}, E.~O. {Ang{\"u}ner} et~al.,
  \href{https://doi.org/10.1051/0004-6361/201732098}{\emph{A\&A} {\bfseries
  612} (2018) A1} [\href{https://arxiv.org/abs/1804.02432}{{\ttfamily
  1804.02432}}].

\bibitem{2015ICRC...34..922Z}
V.~{Zabalza},  in \emph{34th International Cosmic Ray Conference (ICRC2015)},
  vol.~34 of \emph{International Cosmic Ray Conference}, p.~922, July, 2015,
  \href{https://arxiv.org/abs/1509.03319}{{\ttfamily 1509.03319}}.

\bibitem{2022NatAs...6..199L}
R.~{L{\'o}pez-Coto}, E.~{de O{\~n}a Wilhelmi}, F.~{Aharonian}, E.~{Amato} and
  J.~{Hinton}, \href{https://doi.org/10.1038/s41550-021-01580-0}{\emph{Nature
  Astronomy} {\bfseries 6} (2022) 199}
  [\href{https://arxiv.org/abs/2202.06899}{{\ttfamily 2202.06899}}].

\bibitem{Liu_2019}
R.-Y. Liu, C.~Ge, X.-N. Sun and X.-Y. Wang,
  \href{https://doi.org/10.3847/1538-4357/ab125c}{\emph{The Astrophysical
  Journal} {\bfseries 875} (2019) 149}.

\bibitem{2013ApJS..208...17A}
A.~A. {Abdo}, M.~{Ajello}, A.~{Allafort}, L.~{Baldini}, J.~{Ballet},
  G.~{Barbiellini} et~al.,
  \href{https://doi.org/10.1088/0067-0049/208/2/17}{\emph{ApJS} {\bfseries 208}
  (2013) 17} [\href{https://arxiv.org/abs/1305.4385}{{\ttfamily 1305.4385}}].

\bibitem{1978MNRAS.182..147B}
A.~R. {Bell}, \href{https://doi.org/10.1093/mnras/182.2.147}{\emph{MNRAS}
  {\bfseries 182} (1978) 147}.

\bibitem{Berezhko_1999}
E.~G. Berezhko and D.~C. Ellison,
  \href{https://doi.org/10.1086/307993}{\emph{The Astrophysical Journal}
  {\bfseries 526} (1999) 385}.

\bibitem{2012ApJ...761..133Y}
Q.~{Yuan}, S.~{Liu} and X.~{Bi},
  \href{https://doi.org/10.1088/0004-637X/761/2/133}{\emph{ApJ} {\bfseries 761}
  (2012) 133} [\href{https://arxiv.org/abs/1203.0085}{{\ttfamily 1203.0085}}].

\end{thebibliography}\endgroup

\end{document}